\documentclass[%
reprint,
superscriptaddress,
 amsmath,amssymb,
 aps,
prb,
]{revtex4-1}

\usepackage {graphicx,epsfig,graphics,color}
\usepackage{dcolumn}
\usepackage{bm}
\usepackage{hyperref}
\usepackage{sidecap}
\usepackage{float}

\begin{document}
\title{Domain engineering of the metastable domains in the $4f$-uniaxial-ferromagnet CeRu$_2$Ga$_2$B}

\author{D. Wulferding}
\affiliation{Center for Artificial Low Dimensional Electronic Systems, Institute for Basic Science, 77 Cheongam-Ro, Nam-Gu, Pohang 37673, Korea}
\affiliation{Department of Physics, POSTECH, 77 Cheongam-Ro, Nam-Gu, Pohang 37673, Korea}

\author{H. Kim}
\affiliation{Center for Artificial Low Dimensional Electronic Systems, Institute for Basic Science, 77 Cheongam-Ro, Nam-Gu, Pohang 37673, Korea}
\affiliation{Department of Physics, POSTECH, 77 Cheongam-Ro, Nam-Gu, Pohang 37673, Korea}

\author{I. Yang}
\affiliation{Center for Artificial Low Dimensional Electronic Systems, Institute for Basic Science, 77 Cheongam-Ro, Nam-Gu, Pohang 37673, Korea}
\affiliation{Department of Physics, POSTECH, 77 Cheongam-Ro, Nam-Gu, Pohang 37673, Korea}

\author{J. Jeong}
\affiliation{Center for Artificial Low Dimensional Electronic Systems, Institute for Basic Science, 77 Cheongam-Ro, Nam-Gu, Pohang 37673, Korea}
\affiliation{Department of Physics, POSTECH, 77 Cheongam-Ro, Nam-Gu, Pohang 37673, Korea}

\author{K. Barros}
\affiliation{Theoretical Division and CNLS, Los Alamos National Laboratory, Los Alamos, NM 87545, USA}

\author{Y. Kato}
\affiliation{Theoretical Division and CNLS, Los Alamos National Laboratory, Los Alamos, NM 87545, USA}
\affiliation{RIKEN Center for Emergent Matter Science (CEMS), Wako, Saitama 351-0198, Japan}

\author{I. Martin}
\affiliation{Theoretical Division and CNLS, Los Alamos National Laboratory, Los Alamos, NM 87545, USA}
\affiliation{Materials Science Division, Argonne National Laboratory, Argonne, IL 60439, USA}

\author{O. E. Ayala-Valenzuela}
\affiliation{Center for Artificial Low Dimensional Electronic Systems, Institute for Basic Science, 77 Cheongam-Ro, Nam-Gu, Pohang 37673, Korea}
\affiliation{Department of Physics, POSTECH, 77 Cheongam-Ro, Nam-Gu, Pohang 37673, Korea}

\author{M. Lee}
\affiliation{Center for Artificial Low Dimensional Electronic Systems, Institute for Basic Science, 77 Cheongam-Ro, Nam-Gu, Pohang 37673, Korea}
\affiliation{Department of Chemistry, POSTECH, 77 Cheongam-Ro, Nam-Gu, Pohang 37673, Korea}

\author{H. C. Choi}
\affiliation{Center for Artificial Low Dimensional Electronic Systems, Institute for Basic Science, 77 Cheongam-Ro, Nam-Gu, Pohang 37673, Korea}
\affiliation{Department of Chemistry, POSTECH, 77 Cheongam-Ro, Nam-Gu, Pohang 37673, Korea}

\author{F. Ronning}
\affiliation{MPA-CMMS, Los Alamos National Laboratory, Los Alamos, NM 87545, USA}

\author{L. Civale}
\affiliation{MPA-CMMS, Los Alamos National Laboratory, Los Alamos, NM 87545, USA}

\author{R. E. Baumbach}
\affiliation{National High Magnetic Field Laboratory, Tallahassee, FL 32310, USA}

\author{E. D. Bauer}
\affiliation{MPA-CMMS, Los Alamos National Laboratory, Los Alamos, NM 87545, USA}

\author{J. D. Thompson}
\affiliation{MPA-CMMS, Los Alamos National Laboratory, Los Alamos, NM 87545, USA}

\author{R. Movshovich}
\affiliation{MPA-CMMS, Los Alamos National Laboratory, Los Alamos, NM 87545, USA}

\author{J. Kim}
\email[]{Corresponding author: jeehoon@postech.ac.kr}
\affiliation{Center for Artificial Low Dimensional Electronic Systems, Institute for Basic Science, 77 Cheongam-Ro, Nam-Gu, Pohang 37673, Korea}
\affiliation{Department of Physics, POSTECH, 77 Cheongam-Ro, Nam-Gu, Pohang 37673, Korea}

\date{\today}

\begin{abstract}

In search of novel, improved materials for magnetic data storage and spintronic devices, compounds that allow a tailoring of magnetic domain shapes and sizes are essential. Good candidates are materials with intrinsic anisotropies or competing interactions, as they are prone to host various domain phases that can be easily and precisely selected by external tuning parameters such as temperature and magnetic field. Here, we utilize vector magnetic fields to visualize directly the magnetic anisotropy in the uniaxial ferromagnet CeRu$_2$Ga$_2$B. We demonstrate a feasible control both globally and locally of domain shapes and sizes by the external field as well as a smooth transition from single stripe to bubble domains, which opens the door to future applications based on magnetic domain tailoring.

\end{abstract}

\maketitle

\section{Introduction}

The discovery of a skyrmion lattice phase in MnSi~\cite{muhlbauer-09} has been a prime example for self-organized, tunable microscopic magnetic structures with a great prospect in future applications. Subsequent research efforts focused on the non-centrosymmetric, ferromagnetic intermetallics~\cite{skyrmion-yu, skyrmion-munzer, skyrmion-yu-FeGe} and multiferroics.~\cite{seki-12} On the other hand, the observation of skyrmions in a bilayer manganite compound with an inversion center~\cite{yu-14} has shifted the spotlight to the uniaxial ferromagnets, as they allow a tuning of the magnetic behavior based on their shape anisotropy rather than on intrinsic parameters such as the Dzyaloshinskii-Moriya anisotropy.

The recently reported series of Ce-based intermetallics CeRu$_2$$X_2$$M$ ($X$ = Al, Ga; $M$ = B, C) offers an enticing playground to explore various magnetic phases and their respective transitions dictated by the balance between Kondo and Ruderman-Kittel-Kasuya-Yosida (RKKY) interactions~\cite{baumbach-12, baumbach-prb-12}. The title compound CeRu$_2$Ga$_2$B (CRGB) exhibits ferromagnetic order below the Curie temperature of $T_C = 15.4$ K as a result of RKKY interaction magnetic moments of Cerium's localized $4f$ electrons~\cite{baumbach-12}. A strong, Ising-like uniaxial anisotropy has been revealed~\cite{sakai-12}, with the magnetic easy-axis coinciding with the crystallographic $c$ axis, due to the crystalline electric field splitting, which results in the population of the ground state $\Gamma_7^{(1)}$ doublet.~\cite{matsuno-12} While the bulk magnetic properties have been characterized, to some extent, the local domain structure still remains elusive. The influence of vector magnetic fields on the microscopic domain structure as well as the possibility to uncover unusual metastable magnetic phases are interesting, yet open issues for strongly uniaxial ferromagnets.

In this Article, we present the local magnetic state of a 4$f$-uniaxial-ferromagnet CRGB by using a novel magnetic force microscope (MFM) with vector magnetic fields. A periodic modulation of the out-of-plane magnetic moment together with the bubble domain size is directly visualized as the field rotates from the easy axis to the hard axis. This result evidences for the first time an Ising-like domain shape response to the vector field. Our study furthermore demonstrates a tuning of the magnetic domain structure globally via external magnetic field and temperature control, as well as locally via the MFM tip, allowing us to estimate the manipulation force for individual domains. Our experimental observations are supported by theoretical modeling based on time-dependent Ginzburg-Landau dynamics. The highly sensitive, metastable domains observed pave a venue for magnetic memory and spin-torque device applications via magnetic domain engineering.\\

\section{Experimental Details}

Single crystals of CeRu$_2$Ga$_2$B were grown via tri-arc melting synthesis as previously reported~\cite{baumbach-12}. Magnetization measurements were performed in a superconducting quantum interference device (Quantum Design). The dimensions of the sample under investigation are 2.0 mm $\times$ 0.9 mm $\times$ 0.5 mm, resulting in the demagnetization factors $D_a=0.138$ and $D_c=0.547$~\cite{aharoni}. The single crystal was oriented via polarized Raman scattering (WITec alpha 300 R), comparing the intensities of Gallium's $A_{1g}$ and $E_g$ phonon modes as a function of crystal orientation~\cite{drachuck-12}.

MFM measurements were performed in a low temperature MFM system with a home-built MFM probe inside a vector magnet with a field and temperature range of 2-2-9 T (in $x$-$y$-$z$ direction) and 0.3 -- 300 K, respectively~\cite{yang-16}. All experiments were carried out with commercially available MFM tips (PPP-MFMR, Nanosensors). The magnetic force between the tip and the sample results in a frequency shift $\Delta f$ of the tip's resonance frequency $f_0$, which can be related to the force gradient via $\frac{\partial F}{\partial z} = -2k \frac{\Delta f}{f_0}$, where $k$ is the spring constant of the tip.

\section{Magnetic anisotropy}

\begin{figure}
\label{figure1}
\centering
\includegraphics[width=8cm]{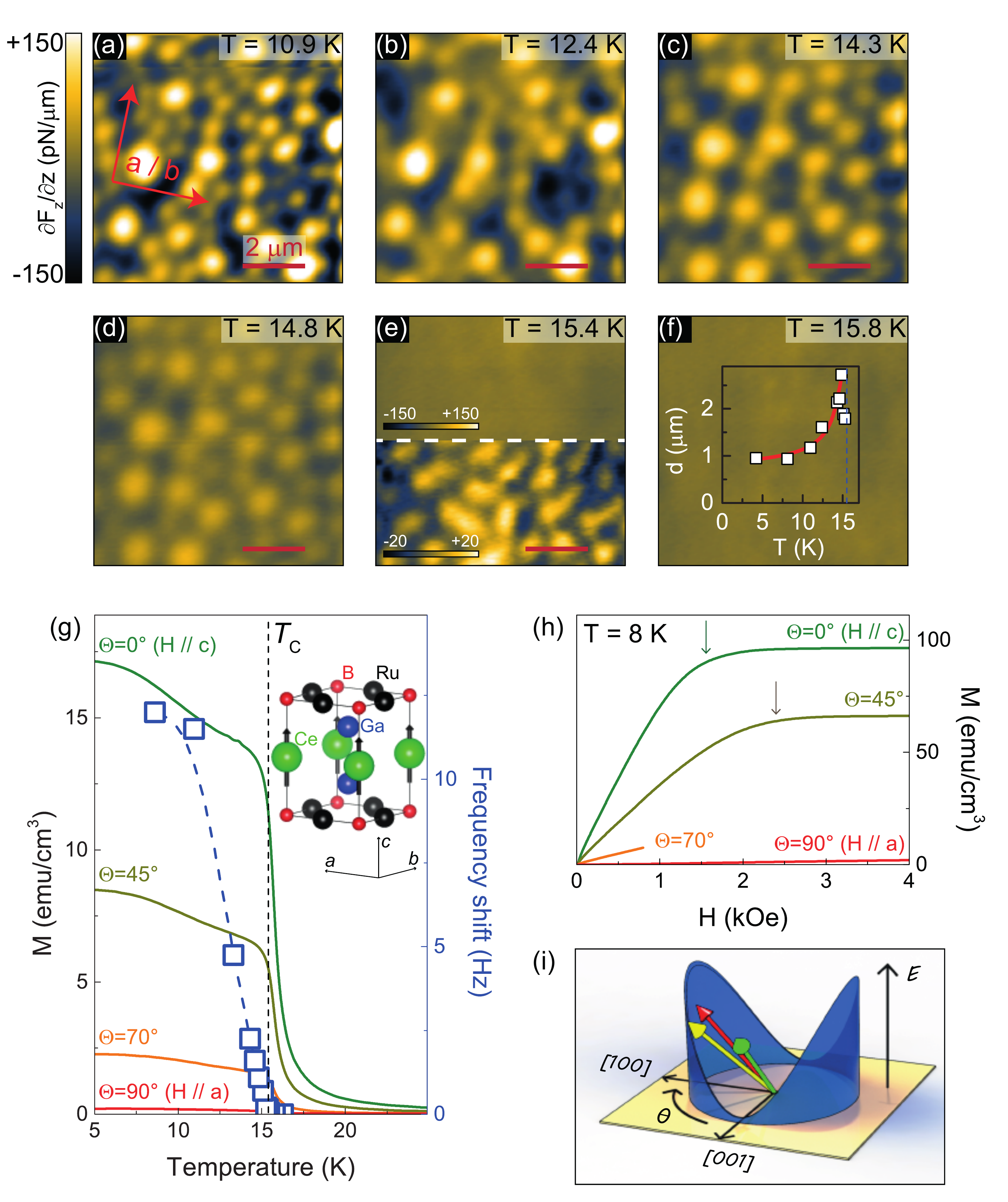}
\caption{\textbf{Magnetic anisotropy in CeRu$_2$Ga$_2$B.} \textbf{a--f}, MFM images obtained with increasing temperatures after initially field-cooling the sample in $H_z = 200$ Oe. The crystallographic orientation for all images is indicated in \textbf{a}. The lower part of \textbf{e} is rescaled in MFM contrast to highlight the domain structure. The data resolution is $128 \times 128$ pixels. The inset in \textbf{f} plots the average bubble size as a function of temperature. \textbf{g}, Maximum frequency shift (open blue squares) obtained from the MFM images together with temperature-dependent magnetization curves at $H_z = 200$ Oe and various field orientations (solid lines). The inset shows the unit cell of CeRu$_2$Ga$_2$B with the spin alignment below $T_C$. \textbf{h}, $M$-$H$ curves for various field orientations measured at $T=8$ K. The arrows denote saturation fields. \textbf{i}, A cartoon illustration of the magnetocrystalline anisotropy energy. The green, yellow, and red arrows correspond to external magnetic fields oriented at $\Theta = 0^{\circ}$, $\Theta = 45^{\circ}$, and $\Theta = 90^{\circ}$, respectively.}
\end{figure}

A basic magnetic characterization of CRGB is given in Fig. 1. The initial bubble state at 10.9 K shown in Fig. 1a is obtained after field-cooling the sample through $T_C$ with $H_z = 200$ Oe. As the temperature approaches $T_C$, the local magnetic moments decrease and the MFM contrast vanishes (see Figs. 1a--f). The comparison between the maximum frequency shift measured in MFM and the magnetization data (open blue squares vs. solid lines in Fig. 1g) reveals a reasonable agreement, considering the fundamental difference between the surface- and bulk sensitive experimental techniques. The averaged bubble size, obtained from a fast Fourier transform (FFT) analysis of the images, starts to increase considerably close to $T_C$, succeeded by a slight decrease approximately 1 K below $T_C$ (see inset in Fig. 1f). The initial increase in bubble diameter $d$ follows a thermally activated behavior of the domain volume, observed in rare earth-transition metal alloys, $d \sim \frac{\delta}{1-T/T_C}$, where $\delta$ corresponds to the domain wall thickness~\cite{labrune-89}. A fit, shown as a thick red solid line in the inset of Fig. 1f, describes our data very well, and yields $\delta \approx 170$ nm. Note that the magnitude of a domain wall thickness $\delta$ may be slightly different from the actual value due to both the tip effect and the imaging condition. In the lower half of Fig. 1e we show the MFM image obtained at 15.4 K rescaled to its full intensity, to emphasize a crossover from round to elongated / distorted bubble domains close to $T_C$. This temperature-driven crossover leads to the decrease of the average domain size. It is related to a reduction of the effective magnetic anisotropy, which becomes comparable to the dipolar interactions close to $T_C$~\cite{choi-07, won-05}. The strong uniaxial anisotropy of CRGB due to the Ising-like spin character (sketched in the inset of Fig. 1g) is evident in the magnetization curves of Fig. 1h. While the saturation magnetization $M_s$ at $T=8$ K is reached around $H=1.5$ kOe for $H // c$ (along the easy axis), it requires slightly higher fields of $H=2.5$ kOe for $\Theta=45^{\circ}$ (see arrows in Fig. 1h) and up to $H\sim 20$ kOe for in-plane fields (not shown). The magnetocrystalline anisotropy energy as a function of field angle is schematically sketched in Fig. 1i. From the $M$-$H$ curve along the hard axis we estimate the first and second order magnetocrystalline anisotropy constants at $T=8$ K as $K_1 = 1.37 \cdot 10^5$ erg/cm$^3$ and $K_2 = 2.10 \cdot 10^5$ erg/cm$^3$ [~\cite{durst-86}]. The total anisotropy energy for a uniaxial system is given by $E_{\mathrm{ani}} = K_0 + K_1 \mathrm{sin}^2 \Theta + K_2 \mathrm{sin}^4 \Theta$ and both $K_1$ and $K_2$ are positive. Therefore, the anisotropy energy will be minimized for out-of-plane magnetic fields ($\Theta = 0^{\circ}$ and $180^{\circ}$)~\cite{cullity}. 

\begin{figure*}
\label{figure2}
\centering
\includegraphics[width=16cm]{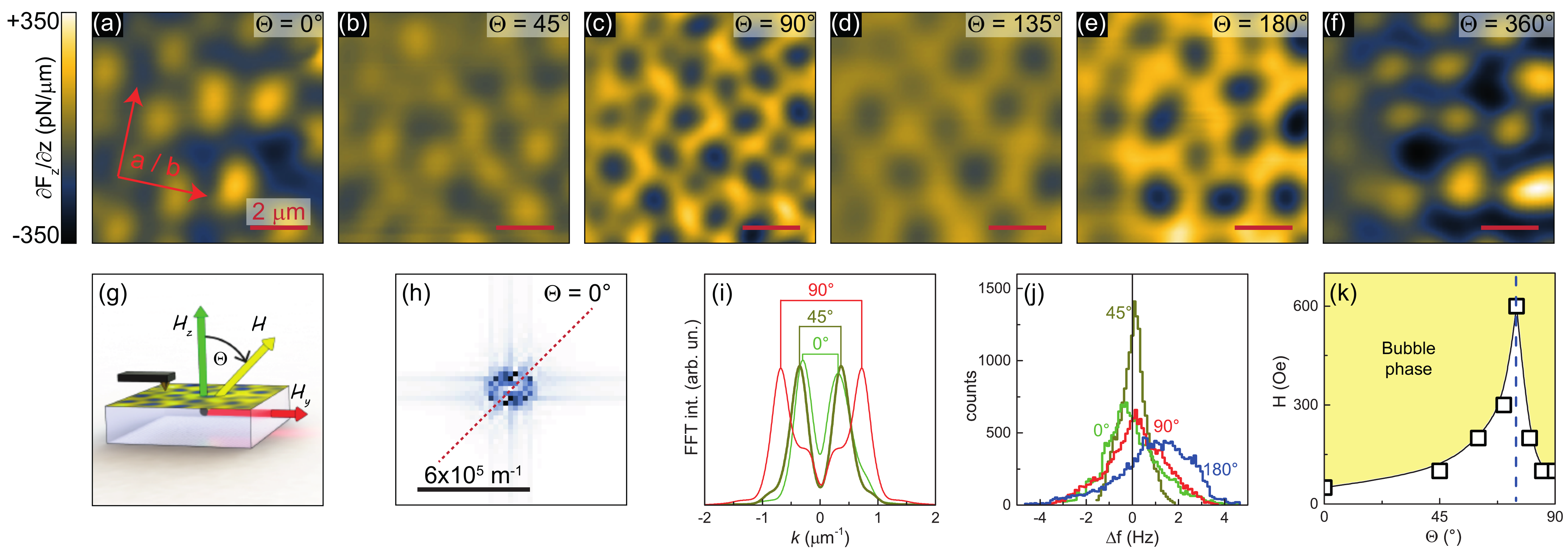}
\caption{\textbf{Domain behavior in vector magnetic fields.} \textbf{a--f}, MFM images obtained at $T=8$ K of CeRu$_2$Ga$_2$B field-cooled with 200 Oe at various field angles $\Theta$. 0$^{\circ}$ (90$^{\circ}$) corresponds to out-of-plane (in-plane) field alignment (see \textbf{g}). The data resolution is $128 \times 128$ pixels. \textbf{h}, Fourier-transformed image of \textbf{a}. \textbf{i}, A comparison of line profiles through Fourier-transformed images at 0$^{\circ}$, 45$^{\circ}$, and 90$^{\circ}$, as indicated by the red dashed line in \textbf{h}. \textbf{j}, Histograms of \textbf{a}, \textbf{b}, \textbf{c}, and \textbf{e}. \textbf{k}, Angle dependent critical field for entering the magnetic bubble domain phase at $T=8$ K.}
\end{figure*}

To understand the impact of this huge intrinsic anisotropy microscopically, we perform MFM experiments in vector magnetic fields. Figs. 2a--f show MFM images of the field-cooled state with $H$= 200 Oe for various field angles $\Theta$, indicated in each image and in Fig. 2g, at 8 K. Field-cooling in the out-of-plane orientation (Fig. 2a) leads to bubble domains with an average diameter of about 1 $\mu$m and a strong signal intensity. Tilting the field by 45$^{\circ}$ towards the plane has no apparent effect on the average bubble size and periodicity, as evidenced by FFT shown in Fig. 2i (brown vs. green curve), where line profiles through the 2-dimensional FFT image are shown (see the red dashed line in Fig. 2h). On the other hand, a pronounced change is observed in the MFM intensity: Bubble domains are of considerably diminished contrast, as seen in Fig. 2b and in the histogram (brown line in Fig. 2j), where the frequency span is rather narrow and the signal centers heavily around $\Delta f = 0$. As an external field of 200 Oe is much smaller than the coercive field of the MFM tip at low temperatures ($\sim 1500$ Oe at 4 K), we rule out a modification of the tip magnetization. We therefore conclude that the out-of-plane component of the magnetic moment decreases due to canted spins, following the direction of the external magnetic field at $\Theta=45^{\circ}$. A closer look at the angular dependence of the MFM contrast reveals that the out-of-plane moment varies in a four-fold fashion -- its full intensity is restored at $\Theta = 90^{\circ}$, $180^{\circ}$, ($270^{\circ}$, not shown) and $360^{\circ}$ ($\widehat{=} 0^{\circ}$), while it is minimized at $45^{\circ}$ and $135^{\circ}$ (as well as $225^{\circ}$ and $315^{\circ}$, not shown). Comparing this behavior with the $M$-$H$ curves in Fig. 1h, we find that aligning spins at 45$^{\circ}$ within the $ac$ plane is relatively easy, while there is an immense energy barrier for rotating spins in-plane (see also cartoon in Fig. 1i). Hence, field-cooling CRGB in a weak in-plane magnetic field ($\Theta = 90^{\circ}$) of 200 Oe leaves the spins mostly aligned along their easy, out-of-plane axis (Fig. 2c). It also causes a zero-net magnetic moment, as neither the ``up''- nor the ``down''-direction is preferred. Note that both average bubble domain size and periodicity are substantially decreased at $\Theta=90^{\circ}$. We can understand this phenomenon by considering the dipolar energy: For magnetic fields along the easy axis ($H \parallel c$) the magnetostatic energy can be partially compensated, resulting in larger domains. Consequently, magnetic fields along the hard axis have no compensating effect and the average domain size will be reduced to minimize the magnetostatic energy~\cite{cullity}. As the external magnetic field continues to rotate back out-of-plane, the domain pattern for $\Theta = 180^{\circ}$ in Fig. 2e is essentially an inverted version of Fig. 2a, indicated by the MFM contrast as well as the histograms (cf. 0$^{\circ}$ with a negative net magnetization and 180$^{\circ}$ with a positive net magnetization in Fig. 2j). Field-cooling the sample in a magnetic field with $\Theta = 360^{\circ}$ (Fig. 2f)  reproduces qualitatively the original MFM image of Fig. 2a. We note, however, that the local distribution of bubble domains has changed, suggesting that the bubble formation is not dominated by a nucleation process around local impurities.

Entering the bubble domain phase requires the presence of a threshold magnetic field (see also Fig. 3). We investigated the critical field strength for entering the bubble domain phase as function of $\Theta$ in Fig. 2k. As we rotate the field from out-of-plane to $\Theta = 75^{\circ}$, the threshold field increases. In contrast to fully out-of-plane fields, tilted fields will partially compensate the uniaxial anisotropy. At the same time, the $z$-component of the field decreases with increasing $\Theta$, thus accounting for the larger threshold-fields. Remarkably, above $\Theta = 75^{\circ}$, this trend is reversed, together with the MFM contrast (cf. Figs. 2a and 2c). This behavior suggests that the magnetocrystalline anisotropy energy increases rapidly for $75^{\circ} \leq \Theta \leq 90^{\circ}$, and becomes too large to be overcome by external magnetic fields of medium strength (i.e., far below saturation). This assumption is supported by the angle dependent magnetization curves in Fig. 1h, where a gradual change in initial slope for $\Theta(0^{\circ} \to 45^{\circ})$ is followed by a more drastic change for $\Theta(45^{\circ} \to 90^{\circ})$. Therefore, large $\Theta$ values can effectively reduce the out-of-plane anisotropy, while inducing a weak in-plane anisotropy. Ultimately, this leads to merging of bright bubbles into maze-like structures, enclosing dark, bubble-like areas, i.e., the observed reversal of the MFM contrast in Fig. 2b vs. Fig. 2c.\\

\begin{figure}
\label{figure3}
\centering
\includegraphics[width=8cm]{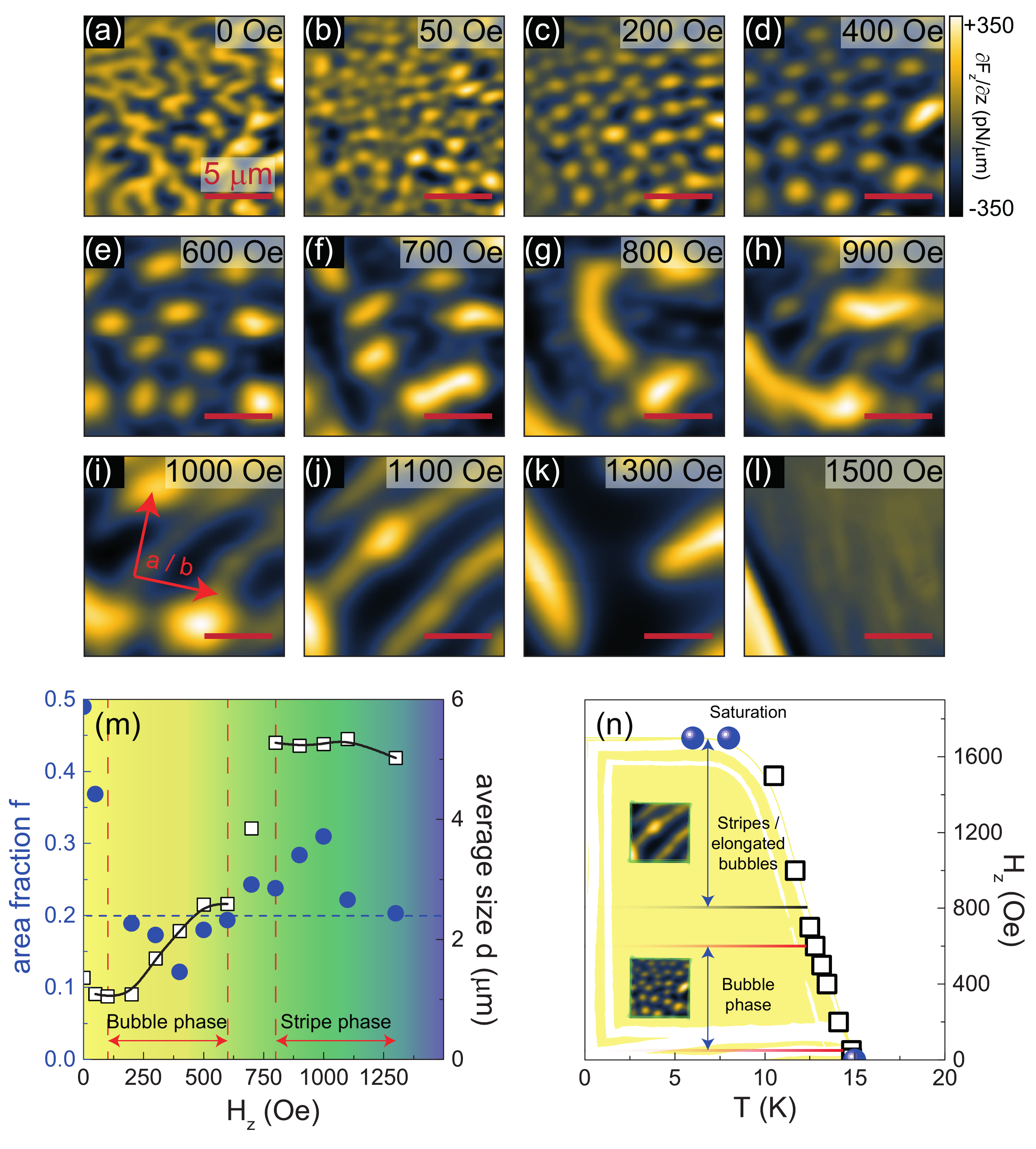}
\caption{\textbf{Evolution of field-cooled magnetic domain structures as a function of out-of-plane magnetic field.} \textbf{a--l}, MFM images at increasing $H_z$; all images were obtained at $T=8$ K. The data resolution is $128 \times 128$ pixels. \textbf{m}, Field dependent area fraction $f$ (blue dots) together with the average size of bubbles and stripes (open squares). \textbf{n}, Magnetic domain phases as a function of magnetic field and temperature. Blue spheres (open boxes): Phase transition as seen in MFM (magnetization) experiments.}
\end{figure}

\section{Domain manipulation}

If we decrease $H$ below the threshold field during the field-cool process, we obtain dendritic domains instead (Fig. 3a). Such a domain structure is characteristic for uniaxial magnets and results from a competition of tendencies to minimize the dipolar energy (by partitioning into smaller domains) and to minimize the domain wall energy (by decreasing the overall length of domain walls and hence forming larger domains)~\cite{goodenough-56, thielsch-12, jeong-15}. After field-cooling in small out-of-plane fields (Figs. 3b and 3c), the additional Zeeman energy results in an imbalance between ``up''- and ``down''-domains, and bubble domains emerge again once a critical area fraction has been reached (see below). Between 600 Oe and 800 Oe the domain pattern changes from bubbles to larger stripes (Figs. 3e--g). The increase in the domain size with increasing magnetic field is counter-intuitive, as one would expect a decrease and subsequent annihilation of minority domains as the Zeeman energy term increases. Instead, we observe a pronounced decrease in the domain wall density. This signals that the energy gain from reducing the number of canted and in-plane spins within the domain walls has to outweigh the energy cost of increasing the area of minority domains. At higher fields the stripes straighten and align at roughly 45$^{\circ}$ from the $a$ and the $b$ axes, i.e., along the [110] and [1$\bar{1}$0] direction (Fig. 3j). This orientational preference is not clearly understood yet, although it is known that single crystalline samples of CRGB suffer from a certain degree of Ga-B intersite mixing~\cite{baumbach-12}. A recent density functional theory study~\cite{hua-12} found that such a replacement can promote ferromagnetic superexchange. Considering the resulting exchange paths of Ce-Ga-Ce are along [110] and [1$\bar{1}$0] would explain the observed preferred stripe direction.
Around 1500 Oe a nearly homogeneous image is obtained, indicating that $H$ is close to saturation (Fig. 3l). A theoretical investigation reveals that a critical area fraction $f_c=\frac{A_{min}}{A_{total}}$ exists (with $A_{min}$ being the area of the minority domains and $A_{total}$ the total area of the MFM frame), that separates the stripe domain phase from the bubble domain phase~\cite{ng-95}. In Fig. 3m we plot the area fraction $f$ as a function of magnetic field $H_z$ (see blue dots). We find that for CRGB the area fraction $f$ in the bubble phase is slightly below 0.2, while it rises above 0.2 in the stripe phase. Therefore, a critical area fraction of $f_c \sim 0.2$ acts as the dividing line, indicated by the dashed blue line. This value is lower than the theoretically predicted $f_c=0.28$ ~[\cite{ng-95}]. On the other hand, perturbations to the theory, such as in-plane anisotropies (e.g., due to intersite mixing), can lead to a reduction of $f_c$~[\cite{ng-95}]. At large fields the area fraction continues to decrease and eventually approaches $f_c$, suggesting the possibility of inducing a bubble domain phase close to saturation (as we shall see in the next section, this is in fact observed). Based on our results we can now construct a $H$-$T$ phase diagram for the different magnetic domains, see Fig. 3n. Bubble domains emerging at small, finite fields transition to elongated bubbles and stripes with increasing $H_z$. The saturation fields obtained from MFM experiments (blue spheres) complement the magnetization data (open white boxes) very well.

In contrast to CRGB, the antiferromagnetically coupled [(Co/Pt)$_8$/Co/Ru]$_{18}$ multilayers~\cite{bran-09, kiselev-10}, and the bilayer manganite La$_{2-2x}$Sr$_{1+2x}$Mn$_2$O$_7$~\cite{jeong-15} with a weak out-of-plane anisotropy both host bubble domain phases only in the vicinity of the magnetic saturation. In the case of CRGB the pre-existing uniaxial anisotropy is already very large. Therefore, $(i)$ the threshold field for entering the bubble phase is relatively smaller, and $(ii)$ the annihilation of minority domains occurs only at high fields. This clearly separates the threshold field for the bubble phase from the magnetic saturation.

\begin{figure}
\label{figure4}
\centering
\includegraphics[width=8cm]{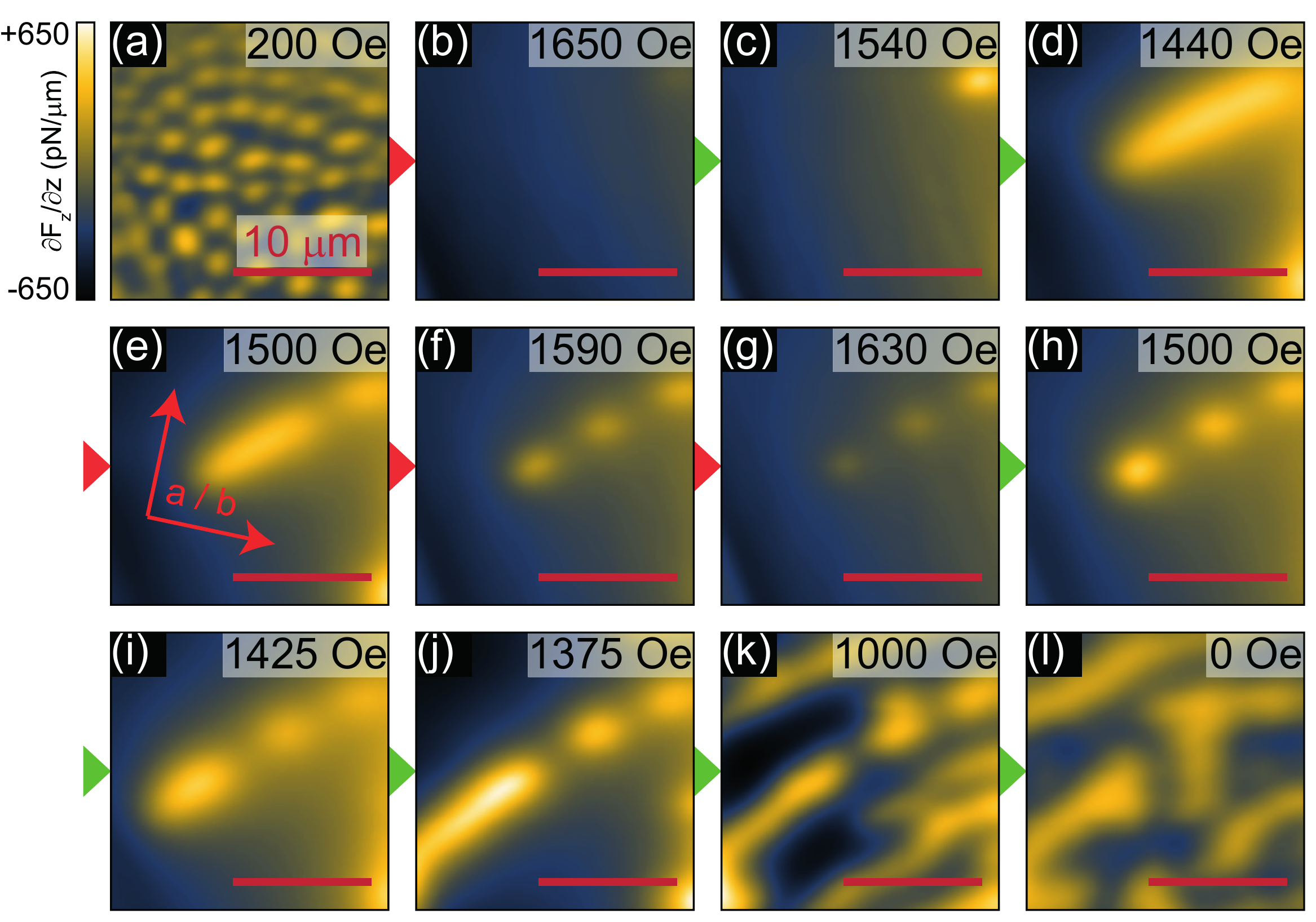}
\caption{\textbf{Domain engineering through field-cycling.} \textbf{a--l}, MFM images of field-cycling experiments at $T=8$ K; the initial field-cooled state with $H_z=200$ Oe is shown in \textbf{a}. Red (green) arrows indicate increasing (decreasing) magnetic fields. The data resolution is $128 \times 128$ pixels.}
\end{figure}

As we have seen so far, the magnetic phases in CRGB decisively depend on the sample's field and temperature history. This behavior can be exploited to tailor magnetic domain structures. Starting from a homogeneous bubble domain phase by field-cooling to 8 K in $H_z = 200$ Oe (Fig. 4a), we increase the magnetic field up to 1650 Oe, which is close to the saturation field (Fig. 4b). Reducing the field down to 1540 Oe leads to the re-occurrence of a single, round shaped magnetic domain in the upper right corner (Fig. 4c). With a further reduction of the magnetic field the domain expands to form a stripe running diagonally across the surface (Fig. 4d), as observed in field-cooled experiments (Fig. 3) and supported by theoretical modeling (see below). In order to check the reversibility, we increase the magnetic field up to 1630 Oe (Figs. 4e--g). Interestingly, the stripe breaks into robust bubbles that only decrease in size as the field approaches saturation. Hence there must exist strong, intrinsic pinning centers that exert little influence on the nucleation of domains while field-cooling. Decreasing the field again from 1630 Oe, the bubble domains gain in size (and hence, in MFM intensity) but they do not merge back into a stripe. Instead, the domain walls remain as barriers, resulting in a broken-up stripe as the magnetic field decreases further (Figs. 4h--j). Meanwhile, the stripe segments widen and start to branch out, while keeping the separating domain wall intact (Figs. 4k--l). The final domain configuration at zero fields resembles the zero-field cooled image in Fig. 3a shape-wise, but the scale is different, with much larger structures after field-cycling. The net magnetization estimated from MFM images is close to zero in both cases, and the $M$-$H$ curves show no significant hysteretic behavior, highlighting that the initial and final states are identical from the bulk point of view. A MFM study of the uniaxial ferromagnet Nd-Fe-B comparing domain structures after thermally demagnetizing and demagnetizing by field reversal found a similar domain size difference~\cite{thielsch-12}. This common observation suggests that systems with a strong uniaxial anisotropy reach their local, microscopic magnetic equilibrium state only after zero-field-cooling, while a demagnetization process via field-cycling leads to a metastable domain state.

\begin{figure}
\label{figure5}
\centering
\includegraphics[width=8cm]{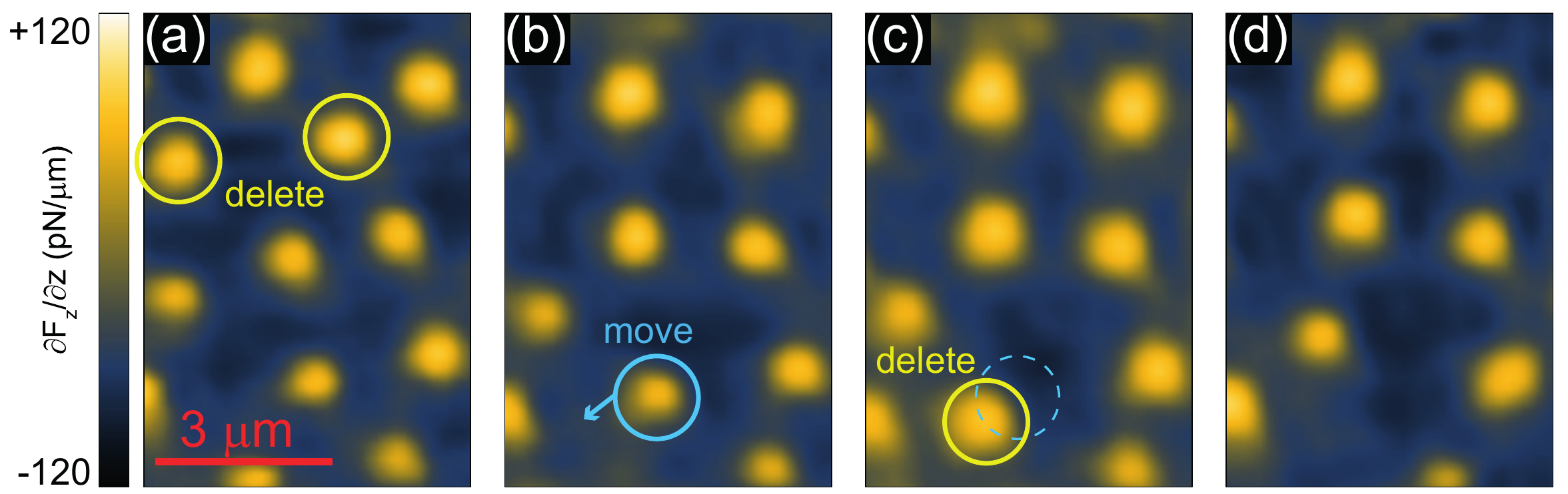}
\caption{\textbf{Local domain manipulation via tip magnetic field.} \textbf{a}, Bubble domains after field-cooling in $H_z = 700$ Oe to $T = 9$ K. \textbf{b}, MFM image after deleting two domains. \textbf{c}, Image after moving the bubble domain marked in \textbf{b}. \textbf{d}, Final MFM image after deleting the moved domain. The data resolution is $48 \times 64$ pixels.}
\end{figure}

While the domain manipulation via vector magnetic fields affects the system globally, we can also manipulate individual bubbles by using the magnetic moment of the tip with a reduced tip-sample distance $d_{\mathrm{ts}}$. In Fig. 5a we create a diluted bubble domain pattern through field-cooling in $H_z = 700$ Oe down to 9 K. The image was obtained with $d_{\mathrm{ts}} = 300$ nm. In order to manipulate the domain pattern, we approach two bubbles consecutively (marked by yellow circles) by the magnetic tip with $d_{\mathrm{ts}} = 20$ nm. Imaging the surface after this procedure with a restored $d_{\mathrm{ts}}$ of 300 nm (Fig. 5b), we notice a successful manipulation, as the bubbles were erased, highlighting opposite directions of magnetization for the bubbles and the tip. In the next step, we position our tip towards the bubble marked by a blue circle ($d_{\mathrm{ts}} = 50$ nm) and move the tip laterally by $\approx 700$ nm, dragging the bubble domain along (Fig. 5c). As a final step, we attempt to delete the dragged bubble via controlled tip-approach (Fig. 5d). Since we know the magnetic moment $m_{\mathrm{tip}}$ of the MFM tip (see supplementary material for details), we can estimate the force necessary to manipulate individual bubbles. We extract the force between the tip and an individual bubble by applying a simple monopole-monopole approximation~\cite{auslaender-09}, $F_{\mathrm{tip-bubble}} = \frac{m_{\mathrm{tip}}\Phi_{\mathrm{bubble}}}{2 \pi}\times \frac{1}{d_{\mathrm{ts}}^2}$, where $\Phi_{\mathrm{bubble}}$ corresponds to the magnetic flux through an individual bubble. We use the signal of a magnetic flux quantum $\Phi_0$ measured on a superconducting Nb film in a comparative experiment to approximate $\Phi_{\mathrm{bubble}} = 4 \times \Phi_0$ (see supplementary material for details). With these values we find that a deleting process takes place through $F_{\mathrm{tip-bubble}} = 13$ nN, while the movement of a bubble requires a force of $F_{\mathrm{tip-bubble}} = 2$ nN, corresponding to an energy cost of $2 \cdot 10^{-15}$ J for a movement over a distance of 1 $\mu$m. The possibility of dragging and deleting single bubble domains in a bulk ferromagnet by the local magnetic tip without a restoring force that would result in a wiggling motion~\cite{auslaender-09} suggests that these domains can be identified as point-like, 0-dimensional objects, rather than 1-dimensional domain strings that penetrate the whole bulk. This assumption is further solidified by the vector magnet experiments (Fig. 2), which did not unveil an elongation of bubbles or a transition into a stripe phase (i.e., the rotation of a cylindrical domain from out-of-plane to in-plane). Instead, bubbles emerge within the $ab$ plane even for fields perpendicular to the $c$ axis. As we will show in the following, theoretical modeling of magnetic domains can well reproduce our experimental observations.

\section{Domain modeling}

\begin{figure}
\label{figure6}
\centering
\includegraphics[width=6cm]{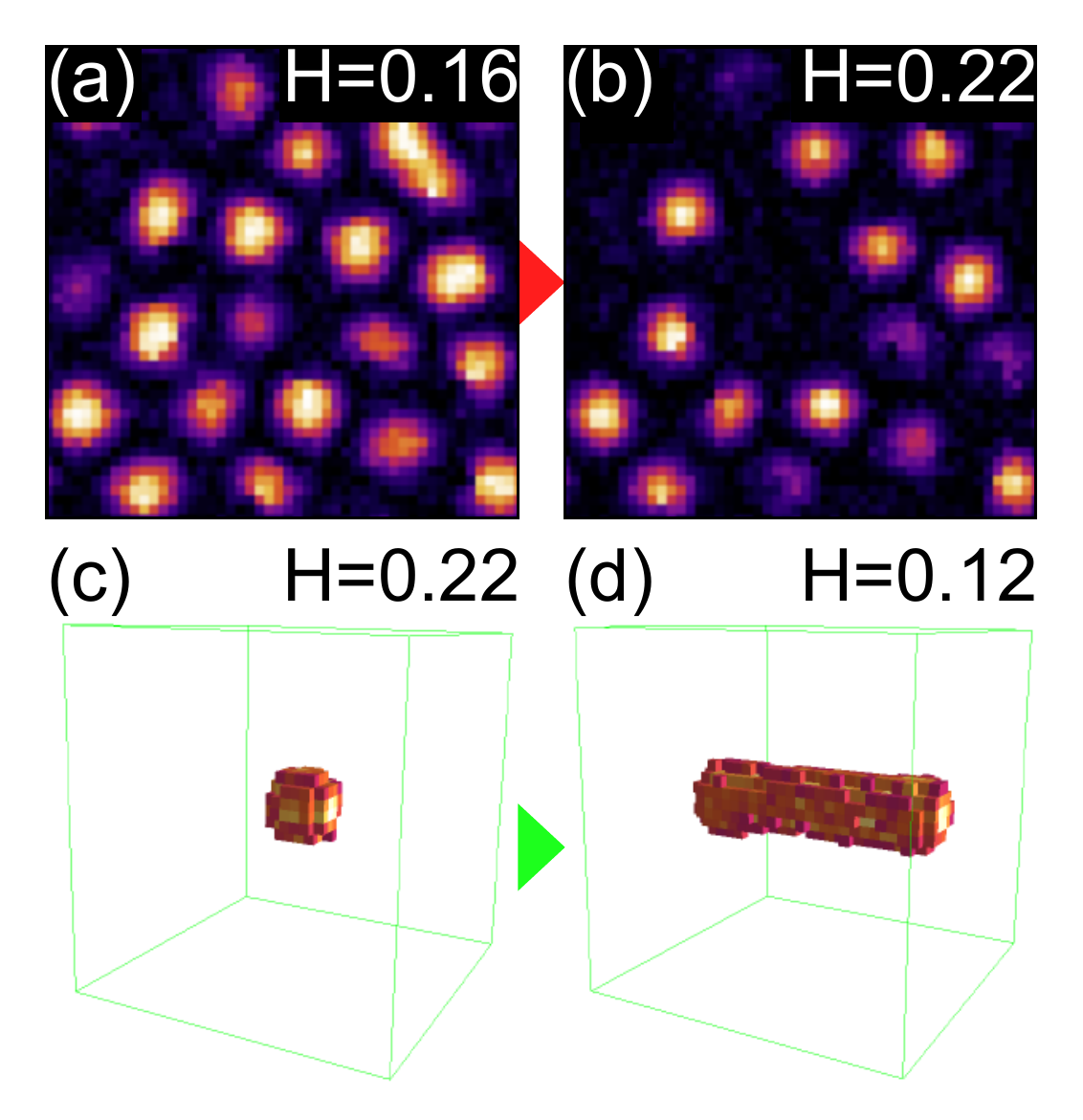}
\caption{\textbf{Domain modeling via Ginzburg-Landau dynamics.} \textbf{a--b}, Decreasing the magnetic bubble density through increasing external magnetic fields. \textbf{c--d}, Domain string elongation after subsequent field decrease, analogous to CeRu$_2$Ga$_2$B.}
\end{figure}

We now turn to a theoretical model of the magnetic system in CRGB to describe the transition among bubbles and stripes. We begin with an effective energy obtained by expanding in powers of the local crystal magnetization $\mathbf S(\mathbf x)$,
\begin{equation}
E[\mathbf{S}]=E_{0}\int\mathrm{d}^{3}\mathbf{x}\left(\frac{1}{2}\mathbf{S}\cdot\Gamma\mathbf{S}-HS_{z}-\frac{1}{2}KS_{z}^{2}\right).
\end{equation}
We work in dimensionless units with $|\mathbf{S}|=1$ and $E_{0}=1$. An external field $H$ points in the $\hat{z}$ direction, $K$ controls the easy-axis anisotropy in $S_{z}$, and $\Gamma$ denotes the interaction operator. To capture the long-wavelength physics, we expand $\Gamma$ in even powers of $\nabla^2$, and truncate to obtain $\Gamma=a^{4}(2q_{0}^{2}\nabla^{2}+\nabla^{4})$, or in Fourier space $\Gamma(k)=a^{4}(-2q_{0}^{2}k^{2}+k^{4})$, with $a$ the lattice constant and $q_{0}$ the preferred spatial frequency. To capture the short-wavelength physics of domain walls, we also apply an ultraviolet cut-off. A large modulation length scale $\lambda = 2 \pi q_{0}^{-1}$ emerges if $q_0^2$ is positive and small. In a 3-dimensional crystal with low carrier density, electron mediated RKKY interactions $\Gamma_{1}=c_{1}a^{2}k^{2}+d_{1}a^{4}k^{4}$ typically favor ferromagnetic ($k = 0$) ordering. If competing interactions (e.g., nearest-neighbor antiferromagnetic superexchange acting on Ce moments) of the form $\Gamma_{2}=-c_{2}a^{2}k^{2}+d_{2}a^{4}k^{4}$ exist, then the \emph{total} interaction $\Gamma = \Gamma_{1} + \Gamma_{2}$ may favor a spatially modulated ($k > 0$) phase. The preferred spatial frequency $q_0 =  \sqrt{(c_{2}-c_{1})/ 2 (d_{1}+d_{2})}/a$ is small if $c_2$ is slightly larger than $c_1$. Empirically, this is what we observe: the bubble domains have a diameter $\lambda \approx 0.5$~$\mu$m much larger than the lattice constant $a = 4.187$~\AA, such that $q_0 a \approx 0.005$. Further modeling details are provided in the supplementary information.

To explore the magnetic domains predicted by our model, we use time-dependent Ginzburg-Landau (TDGL) dynamics, $\partial\mathbf{S}(\mathbf{x},t) / \partial t=-\delta E[\mathbf{S}]/\delta\mathbf{S} +\sqrt{2 k_B T}\eta(\mathbf{x},t)$, where $\eta(\mathbf{x},t)$ is Gaussian white noise. TDGL dynamics is the overdamped limit of physical Landau-Lifshitz-Gilbert dynamics. To construct a dense configuration of bubble domains, we choose dimensionless parameters $K=0.25$, $q_{0}=2\pi/8$, and field cool from $k_B T = \infty$ to $0.01$ at a fixed dimensionless field $H$. In Figs. 6a,b we plot a bulk cross section at $H=0.16$ and $H=0.22$, respectively, colored by the local value of $S_z$, and find good qualitative agreement with the corresponding CRGB experiments (Figs. 3c,d and 4a,b). We note that a finite $K$ is important in stabilizing compact bubbles. At $H \approx 0.22$, bubble annihilation occurs analogous to our experimental observation (Figs. 3d,e). Subsequent \emph{decrease} of the field from $H=0.22$ to $H \approx 0.12$ causes a single bubble to elongate into an extended domain string (Figs. 6c,d), again phenomenologically consistent with our experimental data (Figs. 4c,d).

\section{Summary}

Our comprehensive MFM study highlights the variety of magnetic metastable phases in CeRu$_2$Ga$_2$B and their manipulation as a function of temperature, magnetic field strength and direction. We find the emergence of bubble domains over an extended temperature range as we field-cool the sample in weak out-of-plane magnetic fields. A vector magnet study details the influence of the magnetic field direction on the domain formation in strongly anisotropic ferromagnets. Our field-cycling experiments demonstrate a feasible control over the global domain shape and size, while a magnetic tip achieves a local, selective manipulation of individual bubble domains, opening the door for in-situ domain engineering in uniaxial ferromagnets.\\

\begin{acknowledgments}
This work was supported by the Institute for Basic Science (IBS), Grant No. IBS-R014-D1. Work at Los Alamos was conducted under the auspices of the U.S. Department of Energy, Office of Basic Energy Sciences, Division of Materials Sciences and Engineering.
\end{acknowledgments}

\section{Supplementary}

\subsection{Estimating the manipulation force}

In order to probe the manipulation force of magnetic bubble domains in CeRu$_2$Ga$_2$B, we first estimate the tip magnetic moment per unit length.

\begin{figure}
\label{figure7}
\centering
\includegraphics[width=8cm]{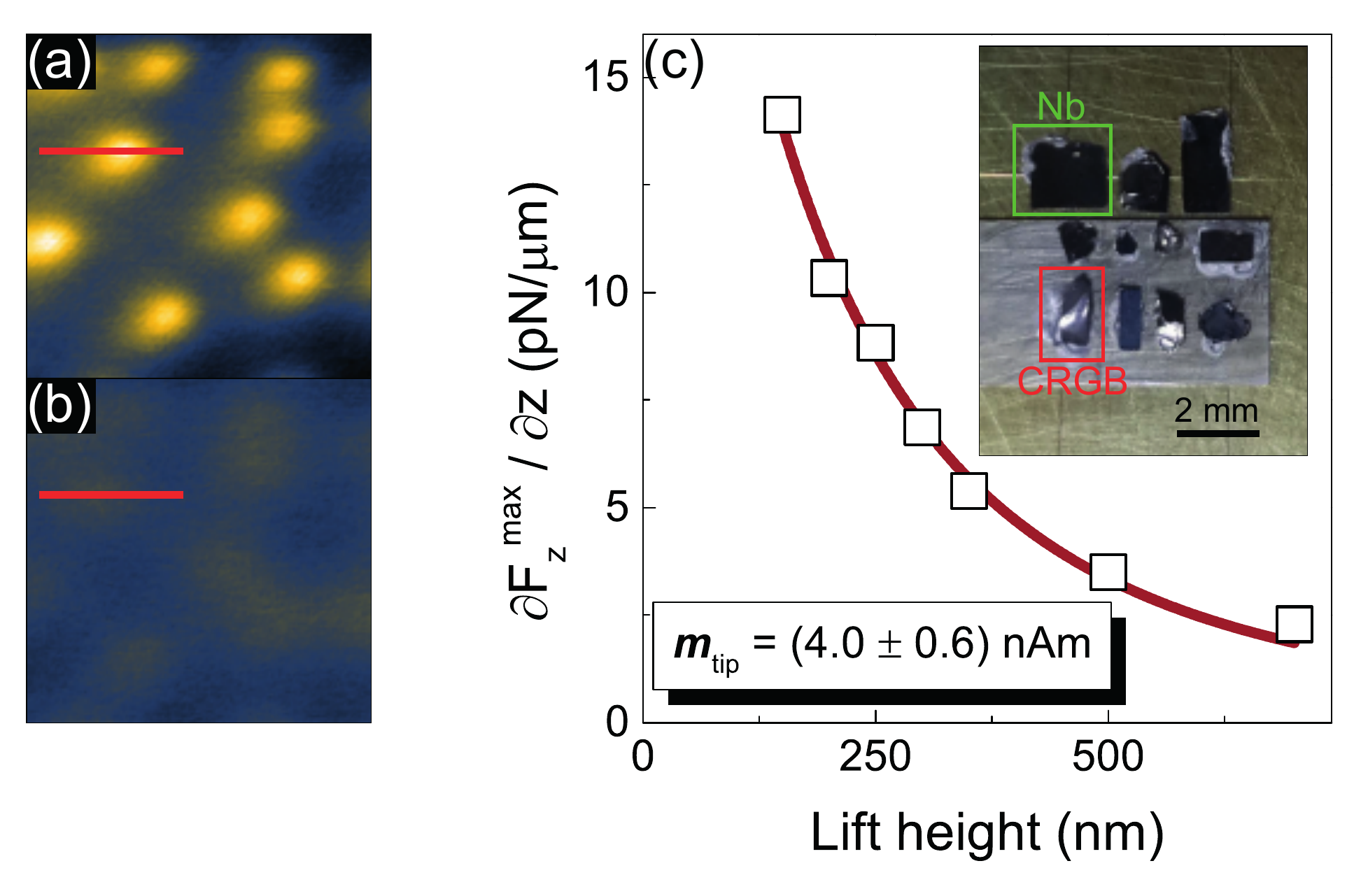}
\caption{Supplement Figure 1: Estimation of the tip magnetic moment per unit length. \textbf{a,b}, MFM images of Abrikosov vortices in a Nb film obtained at $T = 4$ K at a tip-sample distance of 150 nm and 700 nm, respectively. \textbf{c}, Force magnitude as a function of tip-sample distance (open squares). The red solid line is a fit (see text for details). Inset: multi-sample stage, containing various samples investigated during a single cool-down using the same MFM tip.}
\end{figure}

We image Abrikosov vortices in a superconducting Nb film at increasing tip-sample distances, which results in a decrease in MFM contrast (see Figs. S1a,b). Each vortex corresponds to one magnetic flux quantum of $\Phi_0 = h/2e$. Next, we plot the force magnitude $\partial F_z^{max} / \partial z$, extracted from line scans through a single vortex, as a function of the tip-sample distance (see Fig. S1c). In a simple monopole-monopole picture, the interaction between the tip magnetic moment and the point-like vortex can be approximated by~\cite{auslaender-09}
\begin{equation}
\frac{\partial F_z^{max}}{\partial z} = \frac{m_{\mathrm{tip}}\cdot \Phi_0}{\pi} \times \frac{1}{(d_{\mathrm{ts}}+\lambda)^3}
\end{equation}
where $\lambda$ is the London penetration depth ($\approx 110$ nm at $T=4$ K), $d_{\mathrm{ts}}$ is the tip-sample-distance, and $m_{\mathrm{tip}}$ is the tip magnetic moment per unit length. By fitting this approximation to our force gradient data, we obtain $m_{\mathrm{tip}} = (4.0 \pm 0.6)$ nAm (see Fig. S1c).

Next, we estimate the magnetic flux through a single magnetic bubble domain in CRGB by comparing it to a well-defined magnetic flux quantum imaged under the same conditions in superconducting Nb, see Fig. S2.

\begin{figure}
\label{figure8}
\centering
\includegraphics[width=8cm]{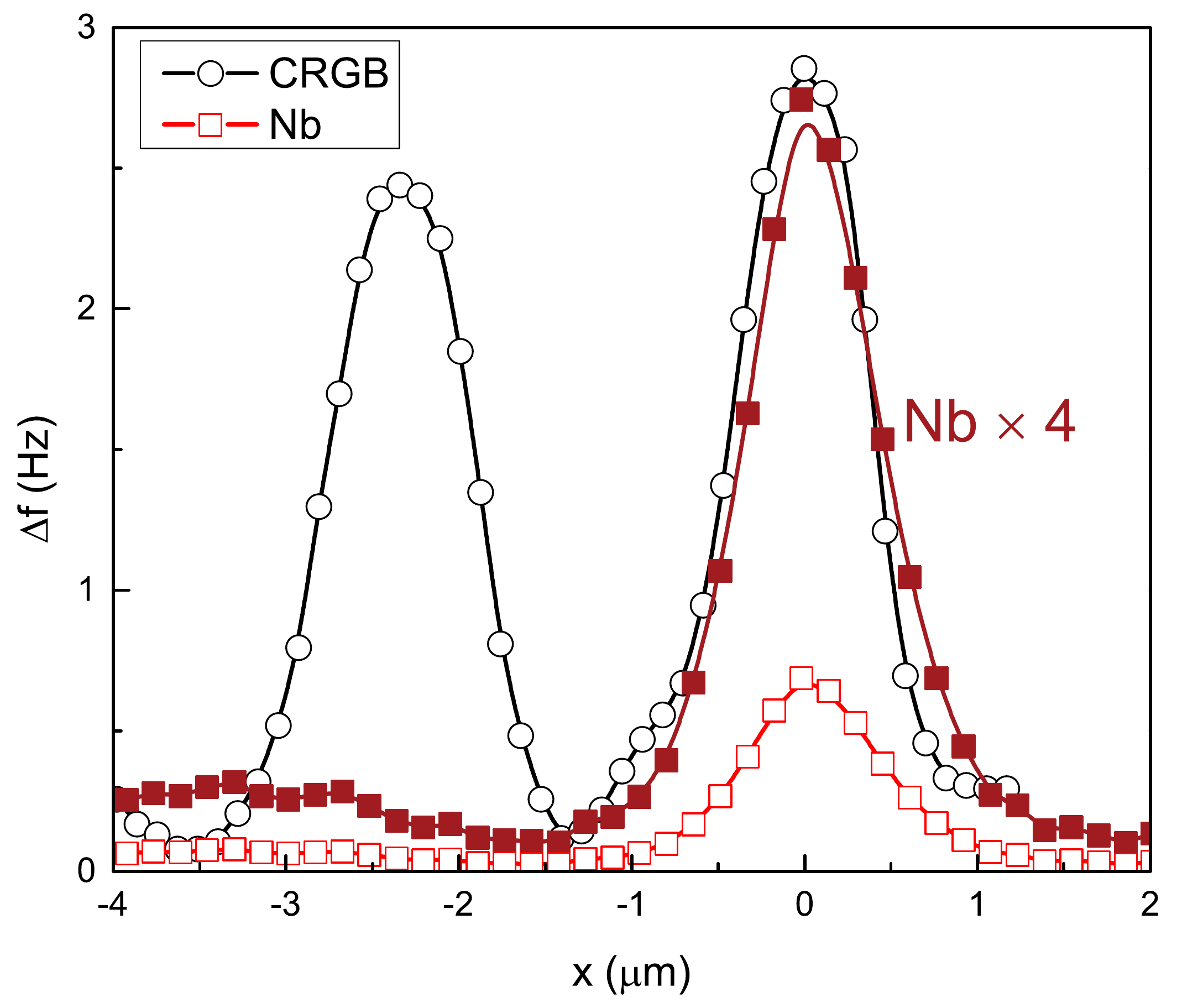}
\caption{Supplement Figure 2: Estimation of the magnetic flux through a single bubble domain. Black curve: line profile through a bubble domain in CRGB imaged at a tip sample distance of 300 nm. Bright red curve: line profile through a magnetic quantum flux imaged in Nb at a tip-sample distance of $(d_{ts} + \lambda) = 300$ nm, dark red curve: Nb vortex profile scaled by a factor of 4.}
\end{figure}

Note that our multi-sample stage~\cite{yang-16} (see inset in Fig. S1c) allows us to investigate both samples, CRGB and the Nb film, during a single cool-down and with the same MFM tip. This feature enables us to perform comparative studies of various samples using a constant, unchanged tip condition. We find that the bubble diameter in CRGB is comparable to the diameter of individual, isolated vortices in Nb. By scaling the intensity of the flux quantum to coincide with the bubble cross-section, we find that a bubble carries a flux $\Phi_{\mathrm{bubble}}$ of roughly 4 times $\Phi_0$. Using these approximations we can now determine the force between tip and bubble necessary for a manipulation process, as described in the main text.

\subsection{Modeling Domain structures in CeRu$_2$Ga$_2$B}

To qualitatively model the appearance of bubbles and their evolution, we use an effective energy functional based on a lowest order Ginzburg-Landau expansion, 
\begin{equation}
E[\mathbf{S}]=\int\mathrm{d}^{3}\mathbf{x}\left(\frac{J}{2}\mathbf{S}\cdot\Gamma\mathbf{S}-HS_{z}-\frac{1}{2}KS_{z}^{2}\right) \;,
\end{equation}
where \textbf{$\mathbf{S}(\mathbf{x})$} is a smooth vector field satisfying $|\mathbf{S}|=1$ (everywhere except bubble cores). We fix energy units by taking the interaction strength to be $J=1$. An external field $H$ points in the $\hat{z}$ direction, and $K$ controls the easy-axis anisotropy in the component $S_{z}$.

The operator $\Gamma$ is most easily understood in Fourier space where the interaction term becomes 
\[
E_{\mathrm{int}}=\frac{1}{2(2\pi)^{-3}}\int\mathrm{d}^{3}\mathbf{k}\Gamma(\mathbf{k})|\mathbf{S}(\mathbf{k})|^{2} \;,
\]
with $\mathbf{S}(\mathbf{k})=\int\mathrm{d}^{3}\mathbf{x}\exp(i\mathbf{k}\cdot\mathbf{x})\mathbf{S}(\mathbf{x})$. In the ferromagnetic Heisenberg model, the choice $\Gamma_{FM}=-\nabla^{2}$ is standard. The representation in Fourier space, $\Gamma_{FM}(\mathbf{k})=|k|^{2}$, has a minimum at $|k|=0$ and thus favors ferromagnetic configurations. Our aim is to model bubbles of some finite size; for that purpose we introduce higher order derivatives to $\Gamma$. The prototypical Swift-Hohenberg model of pattern formation uses $\Gamma_{SH}=(q_{0}^{2}+\nabla^{2})^{2}$ [~\cite{swift-77}], or in Fourier space $\Gamma_{SH}(\mathbf{k})=(q_{0}^{2}-|k|^{2})^{2}$. The minimum of $\Gamma_{SH}$ appears at $|k|=q_{0}$ instead of $0$, which introduces a natural modulation length scale, $\lambda = 2 \pi q_{0}^{-1}$. To model the bubble domains seen in CRGB, we select $\lambda \approx 0.5$~$\mu$m.

\begin{figure}
\label{figure9}
\centering
\includegraphics[width=8cm]{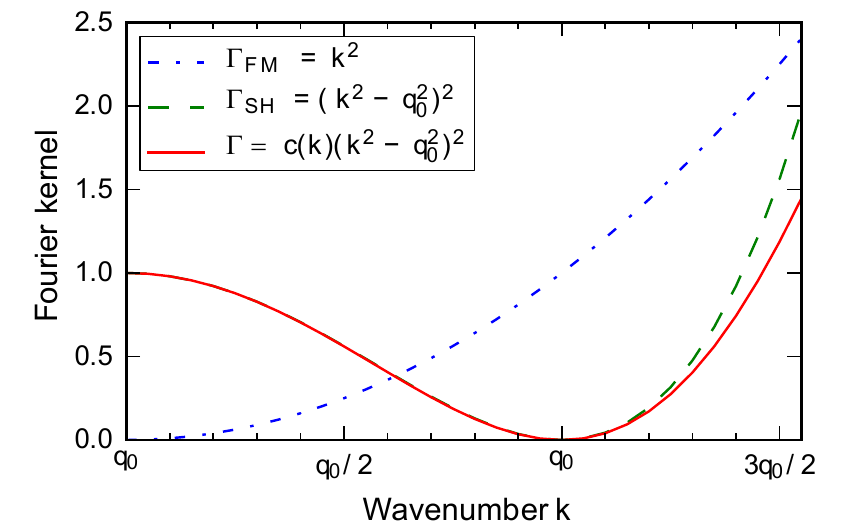}
\caption{Supplement Figure 3: Possible Fourier kernels for the interaction term. The choice $\Gamma_{FM}=k^{2}$ corresponds to the Heisenberg ferromagnet. The choice $\Gamma_{SH}=(q_{0}^{2}-k^{2})^{2}$ corresponds to the Swift-Hohenberg model, and induces modulations on the scale $\lambda\sim1/q_{0}$. We use $\Gamma_{SH}$ with a damping factor $c(k)=(1+k^{4}/(2q_{0})^{4})^{-1}$ to eliminate UV divergences while preserving the physics at $k\sim q_{0}$.}
\end{figure}

The model above is valid at large length scales, and assumes a smooth field $\mathbf S(\mathbf x)$. To represent bubbles as finite energy defects, our model has to be regularized in the small wavelength, ultra-violet (UV) limit. The
core of a bubble singularity may be represented as $\mathbf{S}=\mathbf{r}/|r|$. We can estimate $E_{\mathrm{core}}$ for a single bubble using dimensional analysis. The large $k$ scaling of the Swift-Hohenberg operator, $\Gamma_{SH}\sim k^4$, causes the bubble core energy to diverge like the UV cutoff frequency, $\Lambda$. The physical lattice constant $a$ (order of 1 nanometer) is not a good regulator, because selecting $\Lambda\sim1/a$ would lead to a bubble core energy on the order of $1/a$, which is very large compared to the characteristic non-singular part of the configuration energy, $\sim q_{0}$. In any case, the UV divergence is not physical since there is no reason to believe that $\Gamma(\mathbf{k})\sim k^{4}$ at large $|k|$. We regularize the model at large $k$ by choosing
\begin{eqnarray}
\Gamma(\mathbf{k}) & = & c(|k|)(q_{0}^{2}-|k|^{2})^{2}\label{eq:gamma}\\
c(k) & = & \frac{1}{1+k^{4}/\Lambda^{4}}
\end{eqnarray}
With this choice, $\Gamma\sim1$ at frequencies $k\gg \Lambda$. The frequency cutoff $\Lambda$ controls the bubble core energy. We choose $\Lambda=2q_{0}$, where $q_{0}$ is the modulation frequency. For this choice, the bubble energy is comparable to the energy of the entire bubble configuration. As shown in Fig. S3, our choice of $\Lambda=2q_{0}$ is sufficiently large that $\Gamma$ is nearly unaffected at its minimum at $k\sim q_{0}$.

To summarize our theoretical model, it is a minimal one that contains just the four necessary physical parameters: (1) external field $H$, (2) anisotropy $K$, (3) modulation frequency $q_{0}$, and (4) finite bubble core energy via the frequency cutoff $\Lambda$. An alternative regularization of the bubble core energy is to ``soften'' the spins by lifting the restriction $|\mathbf{S}|=1$.

The form of the interaction that we constructed is essentially phenomenological, and motivated by the CRGB data. However, we illustrate one scenario of how such an interaction may come about. Consider that the interaction between Ce moments in CRGB is due to two primary interactions: nearest neighbor antiferromagnetic superexchange~\cite{anderson-50} and longer-range oscillatory in space RKKY interaction~\cite{ruderman-54, kasuya-56, yosida-57}. Let us show that combining these two interactions one can naturally obtain a long-wavelength modulation.
In momentum space, 
\begin{eqnarray}
J_{AF}(k)&=& J_0[\cos(k_x a)+\cos(k_y a)+\cos(k_z a)] \\
J_{RKKY}(k)&=& -J_1\chi(k), 
\end{eqnarray} 
where $\chi(q)$ is the Lindhard function~\cite{lindhard-54}. Expanding around $k=0$ and dropping an irrelevant constant term, we obtain $J_{tot} = A\left[(k/2k_F)^2 + (k/2k_F)^4/5+ \ldots\right] + B\left[-(ka)^2 + (k_x^4 + k_y^4 + k_z^4)/12+\ldots\right]$.
By tuning parameters $A$ and $B$, we can cancel the $k^2$ terms leaving $J\propto k^4/(2k_F)^2/5 +(k_x^4 + k_y^4 + k_z^4)a^2/12$. Allowing the quadratic AF term to slightly dominate, and taking $k_Fa\ll1$, we obtain the desired nearly isotropic form of the interaction peaked at a finite but small value of $k$.

To perform simulations on this model, we evolve the field $\mathbf{S}(\mathbf{x},t)$ according to the Langevin equation 
\begin{equation}
\frac{\partial\mathbf{S}}{\partial t}=-\frac{\delta E}{\delta\mathbf{S}}+\sqrt{2k_B T}\eta,  \label{eq:lang_dyn}
\end{equation}
where $k_{B}T$ is the temperature in scaled energy units and $\eta(\mathbf{x},t)$ is Gaussian white noise with moments $\langle\eta(\mathbf{x},t)\rangle=0$ and $\langle\eta(\mathbf{x},t)\eta(\mathbf{x}',t')\rangle=\delta(\mathbf{x}-\mathbf{x}')\delta(t-t')$. The functional derivative of energy is 
\begin{equation}
\frac{\delta E}{\delta\mathbf{S}}=\int\mathrm{d^{3}\mathbf{x}}\Gamma\mathbf{S}(\mathbf{x})-H\hat{z}-KS_{z}\hat{z}.
\end{equation}
To satisfy the constraint $|\mathbf{S}|=1$, we modify the time evolution with a Lagrange multiplier term $\partial\mathbf{S}/\partial t\rightarrow\partial\mathbf{S}/\partial t+\lambda\mathbf{S}$, where $\lambda$ is to be solved self-consistently. Assuming ergodicity, this Langevin equation generates fields $\mathbf{S}$ with the appropriate Boltzmann distribution, $P[\mathbf{S}]\propto\exp(-\beta E)$. In the spirit of time-dependent Ginzburg-Landau modeling, we will interpret the evolution of $\mathbf{S}$ as a qualitative description of the \emph{non-equilibrium} dynamical evolution of $\mathbf{S}$.

We integrate the Langevin equation using an implicit numerical scheme for stability, and alternating between Fourier and real space for efficiency. The steps are: 
\begin{enumerate}
\item Use the Fast Fourier Transform (FFT) to calculate $\mathbf{S}(\mathbf{k},t)$. In Fourier space, construct $\mathbf{T}(\mathbf{k},t)\equiv(1+\Delta t\Gamma(\mathbf{k}))\mathbf{S}(\mathbf{k},t)$,
using the functional form for $\Gamma(\mathbf{k})$ given in Eq.~\eqref{eq:gamma}. 
\item Use a reverse FFT to calculate $\mathbf{T}(\mathbf{x},t)$. 
\item Update $\mathbf{S}(x,t)$ according to 
\[
\mathbf{S}(\mathbf{x},t+\Delta t)=\mathbf{T}(\mathbf{x},t)+\Delta t\hat{z}(H+KS_{z})+\sqrt{2 k_B T \Delta t}\xi_{\mathbf{x},t}
\]
where each $\xi_{\mathbf{x},t}$ is a Gaussian random number with unit variance. 
\item Normalize $\mathbf{S}(\mathbf{x},t+\Delta t)$ to satisfy the constraint
$|\mathbf{S}|=1$. 
\end{enumerate}

\begin{figure}
\label{figure10}
\centering
\includegraphics[width=8cm]{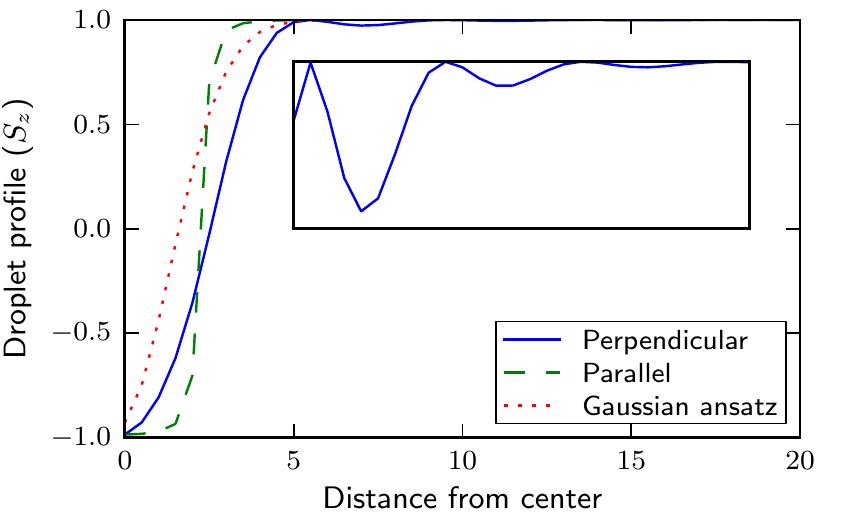}
\caption{Supplement Figure 4: The profile of $S_{z}$ as a function of distance from the bubble center. The profile decays quickly in both parallel and perpendicular directions, consistent with the numerically observed weak bubble-bubble interaction. The parallel direction, passing through the bubble defect, has a sharp jump in $S_{z}$ at $r\approx2$ at the singularity. A limitation of our Gaussian ansatz is that it is radially symmetric, and does not contain a sharp jump in $S_z$. Inset: Zoom-in around $S_z=+1.0$.}
\end{figure}

\begin{figure}
\label{figure11}
\centering
\includegraphics[width=8cm]{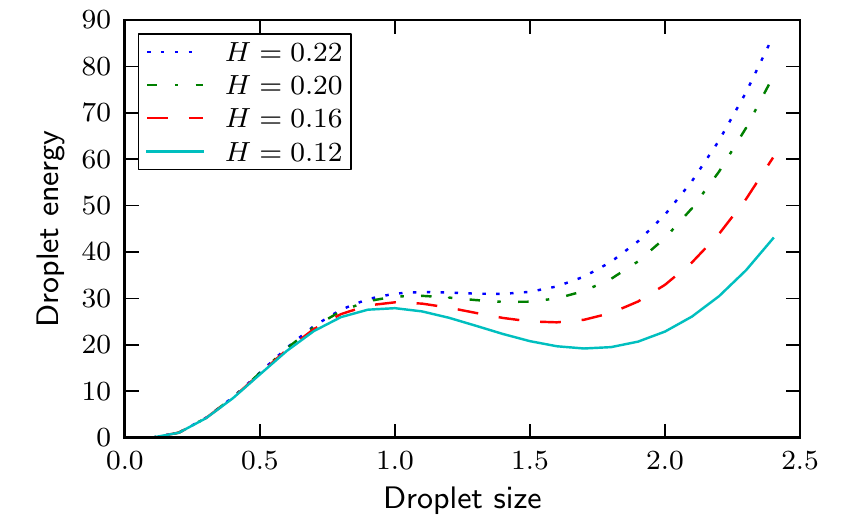}
\caption{Supplement Figure 5: Energy of a single bubble as a function of linear size, using the Gaussian ansatz. A well defined energetic minimum indicates preferred bubble size. At fields $H\gtrsim0.22$ the bubble is unstable to decreasing size $\sigma$ and dynamically implodes. At fields $H\lesssim0.12$ the bubble is unstable to deconfinement.}
\end{figure}

The bubbles we observe are locally stable, but do not globally minimize the energy. For these system parameters, the true energy minimum is actually the ferromagnetic state $\mathbf{S}\approx+\hat{z}$. To understand the energetics of a single bubble, we apply the ansatz,
\begin{eqnarray*}
S_{z} & = & 1-2e^{-r^{2}/2\sigma^{2}}\\
S_{x} & = & \cos\phi\sqrt{1-S_{z}^{2}}\\
S_{y} & = & \sin\phi\sqrt{1-S_{z}^{2}}
\end{eqnarray*}
where $r$ is the radial distance from the origin and $\phi$ is the azimuthal angle. Our Gaussian ansatz has a single parameter $\sigma$, the linear bubble size. For this ansatz, $S_{z}=0$ at radial distance $r_{0}=\sigma\sqrt{2\ln2}\approx1.18\sigma$. Our ansatz does not capture the bubble defects precisely. Instead, the entire $z$ axis is singular because $S_{x}$ and $S_{y}$ are ill-defined on this line. In an actual bubble of minimal energy, we would observe $S_{z}=-1$ along the entire $z$ axis when $r<r_{0}$, and $S_{z}=+1$ when $r>r_{0}$. This discrepancy is illustrated in Fig. S4, where a real bubble of minimum energy is compared to the Gaussian
ansatz.

The Gaussian ansatz correctly predicts bubble annihilation at large external field $H$. In Fig. S5 we plot the ansatz energy as a function of bubble size $\sigma$ at fixed $K=0.25$ and varying $H$. The pronounced energetic minimum indicates a preferred bubble size that decreases slightly with increasing field $H$. At fields $H\gtrsim0.22$ the local minimum is removed and the bubble annihilates, leaving behind pure ferromagnetic alignment. This behavior is consistent with what we observe in Langevin dynamics simulations, Eq.~\eqref{eq:lang_dyn}. Our Langevin simulations also demonstrate that the bubble is unstable to deconfinement (and extension of the string) when $H\lesssim0.12$.

\end{document}